\documentclass[12pt,a4paper]{article}

\usepackage[scale={.75,.8}]{geometry} 
\usepackage{pstricks}
\usepackage[dvips]{graphicx}
\usepackage[latin2]{inputenc}
\usepackage{epic}
\usepackage{latexsym}
\usepackage{epsf}
\usepackage{epsfig}
\usepackage{amssymb,amsmath}
\usepackage{cite} 

\newcommand{\fref}[1]{fig.\ \ref{f.#1}} 
\newcommand{\eref}[1]{eq.\ (\ref{e.#1})} 
\newcommand{\erefn}[1]{ (\ref{e.#1})}

\newcommand{\sref}[1]{Section \ref{s.#1}}
\newcommand{\cref}[1]{Chapter \ref{c.#1}}

\def\nn{\nonumber \\}  
\newcommand{\nl}{& \nonumber \\ &}
\newcommand{\bnl}{\right . & \nonumber \\ & \left .}

\def\ds{\displaystyle}

\def\beq{\begin{equation}} 
\def\eeq{\end{equation}} 
\def\bea{\begin{eqnarray}}  
\def\eea{\end{eqnarray}}  
\newcommand{\bal}{\begin{align}}
\newcommand{\eal}{\end{align}}   
\def\ba{\begin{array}}  
\def\ea{\end{array}}   
\def\bi{\begin{itemize}}  
\def\ei{\end{itemize}}  
\def\ben{\begin{enumerate}}  
\def\een{\end{enumerate}}  
\def\beq{\begin{equation}}  
\def\eeq{\end{equation}}  
\def\bc{\begin{center}}
\def\ec{\end{center}} 
 \def\bt{\begin{table}}
\def\et{\end{table}}  
 \def\btb{\begin{tabular}}
\def\etb{\end{tabular}}

\def\cm{{\mathcal M}}  
\def\co{{\mathcal O}}

\def\gev{\, {\rm GeV}}

\def\mass2{mass${}^2$}

\newcommand{\mkk}{M_{\mathrm{KK}}}
\newcommand{\mpl}{M_{\mathrm{Pl}}}
\newcommand{\gsm}{G_{\mathrm{SM}}}

\def\ads{${\mathrm A \mathrm d \mathrm S}_5$}

\def\ra{\rangle}
\def\la{\langle}  

\def\pa{\partial}

\newcommand{\tr}{\rm Tr}

\def\simgt{\stackrel{>}{{}_\sim}}

\newcommand{\ha}{{\hat a}}
\newcommand{\hab}{{\hat b}}
\newcommand{\hac}{{\hat c}}
\newcommand{\ham}{{\hat m}}
\newcommand{\han}{{\hat n}}

\newcommand{\ti}{\tilde}

\def\ov{\overline}

  \def\eps{\epsilon}
\begin{document}

\pagestyle{empty}
\begin{flushright}
CERN-PH-TH/2006-217\\ 

{\bf \today}
\end{flushright}
\vspace*{5mm}
\begin{center}

\renewcommand{\thefootnote}{\fnsymbol{footnote}}

{\large {\bf About the holographic pseudo-Goldstone boson}} \\ 
\vspace*{1cm}
{\bf Adam~Falkowski}\footnote{Email:adam.falkowski@cern.ch}

\vspace{0.5cm}

CERN Theory Division, CH-1211 Geneva 23, Switzerland\\
Institute of Theoretical Physics, Warsaw University, \\ Ho\.za 69, 00-681 Warsaw, Poland

\vspace*{1.7cm}
{\bf Abstract}
\end{center}
\vspace*{5mm}
\noindent 
Pseudo-Goldstone bosons in 4D strongly coupled theories have a dual description in terms of 5D gauge theories in warped backgrounds.    
We introduce systematic methods of computing the pseudo-Goldstone potential for an arbitrary warp factor in 5D. 
When applied to electroweak symmetry breaking, our approach clarifies the relation of physical observables to geometrical quantities in five dimensions.   

\vspace*{1.0cm}
\date{\today}


\vspace*{0.2cm}
 
\vfill\eject
\newpage

\setcounter{page}{1}
\pagestyle{plain}

\renewcommand{\thefootnote}{\arabic{footnote}}
\setcounter{footnote}{0}

\section{Introduction}

The LHC experiment will soon probe how $W$ and $Z$ bosons acquire their  masses.  
So far we have only indirect hints about the actual mechanism of  
electroweak symmetry breaking. 
The Standard Model (SM) introduces a weakly coupled fundamental scalar field --- the Higgs field --- whose condensation breaks the electroweak symmetry.   
This simple picture is consistent with  precision electroweak measurements, if the mass of the Higgs boson
 is $\simgt 100 \gev$ 
and the SM cutoff is far above a TeV.  
However, from a theoretical point of view, a fundamental scalar much lighter than the theory's cutoff is unnatural, unless there is a symmetry protecting its mass.

Perhaps the most elegant realization of the Higgs sector is  the one in which the Higgs field is composite, rather than fundamental \cite{georgi}. 
This can be achieved by introducing a  new  strongly interacting sector with a global symmetry 
$G \supset \gsm$. 
In this scenario the Higgs field is a pseudo-Goldstone boson arising from a  spontaneous breaking of $G$.  
The global symmetry could be broken by a condensate, in close analogy with chiral symmetry breaking in QCD. 
If the condensate leaves the SM gauge symmetry intact, the effective theory below the compositeness scale of the new strong interactions includes the SM fields coupled to a non-linear sigma model field describing the pseudo-Goldstone Higgs. 
The Higgs field has a flat potential at tree level, but a non-trivial potential can be generated through its couplings to the SM that do not respect the full global symmetry.
In such framework, we expect the Higgs potential of the form  
\beq
V(h) \sim {1 \over 4 \pi^2} \int_0^\infty dp \, p^3 \log (1 + F(p^2) \sin^2(h/f_h) ) \, . 
\eeq
Here $f_h$ is the scale of the spontaneous global symmetry breaking. 
The potential depends on the Higgs via $\sin(h/f_h)$ because Goldstone fields are  described by periodic variables.   
The form factor $F(p^2)$ depends on  the effective theory interactions between the SM gauge and fermion fields and the pseudo-Goldstone Higgs.

The main drawback of this scenario is that strongly coupled theories are notoriously difficult to master. 
Typically, the form factors can only be estimated by naive dimensional analysis.  
However, it has recently been  realized that at least some 4D strongly coupled sectors --- those with a large number of colours of the strong gauge group --- can be represented by perturbative 5D gauge theories. 
The 5D  gravitational background is curved with the line element $ds^2 = a^2(x_5)(dx)^2 - dx_5^2$ and the 5th dimension is truncated by a UV brane at $x_5 = 0$ and by an IR brane at $x_5 = L$. 
This is an example of holography \cite{adscft} applied to 5D  field theories \cite{adscftrs}. 
According to the holographic dictionary, the 5D gauge symmetry $G$ broken by boundary conditions on the IR brane  corresponds to the spontaneously broken global symmetry $G$.
Perturbative expansion in the 5D gauge coupling $g_5$ corresponds to $1/N_c$ expansion in 4D. 
On the 5D side, the spectrum contains a scalar excitation originating from  the 5th component of the gauge field along the broken gauge group generators.
This excitation can be identified with the 4D pseudo-Goldstone boson. 
Next, the 5D gauge symmetry is reduced on the UV brane, which corresponds to explicit breaking of the global symmetry in 4D  by couplings of external fields to the strongly interacting sector.
The explicit breaking generates a potential  for the pseudo-Goldstone Higgs.  
The important advantage of the 5D approach is  that the Higgs potential can be computed perturbatively.  

In fact,  computations of the pseudo-Goldstone Higgs potential in the 5D framework were first performed  independently of the holographic interpretation \cite{ho}.
Instead, the motivation was the identification of the Higgs field with a component of a gauge field. 
This gauge--Higgs unification provides a solution to the hierarchy problem from the 5D  perspective. 
Indeed, loop corrections to the Higgs potential  are necessarily finite; the divergent corrections are not allowed because the gauge-invariant operator that can yield a Higgs mass is non-local in the 5th coordinate.

At the phenomenological level, most of the activity so far has concentrated on gauge--Higgs unification in the flat Minkowski background \cite{flat}, which does not admit a holographic interpretation. 
Generalization to the \ads \, background (which corresponds to conformal  symmetry of the strongly coupled sector) was accomplished quite recently, in \cite{conopo,odwe,agcopo,homa,honosa} and the holographic aspect was stressed in  \cite{conopo,agcopo}.
The models in \ads \, have reached the stage of fully realistic extensions of the SM, as shown by the calculation of electroweak observables in \cite{agco}.

In this paper we advocate methods that allow the computation of  the form factors entering the pseudo-Goldstone potential for an arbitrary 5D warped background. 
This generalization corresponds to deviations of the  4D strong sector  from conformal symmetry.
Such deviations are, in fact, expected in realistic models of the strong sector,  in which the hierarchy of scales between the UV brane and the IR brane is stabilized \cite{gowi}.  
We show that the form factors  can be simply expressed in terms of the solutions to the 5D equations of motion. 
Moreover, in many cases the form factors can be adequately approximated by a simple analytic expression: 
\beq
F(-p^2) \approx  {y^2 f_h^2 \over \mkk^2} {1 \over \sinh^2(p/\mkk) }  \, , 
\eeq
where $y$ stands for a low-energy Yukawa or gauge coupling. 
The Kaluza--Klein (KK) scale $\mkk$ is roughly equal the mass of the lightest KK mode 
and corresponds to the compositeness scale in 4D. 
At low energies, below the KK/compositeness scale, the form factor  exhibits the $1/p^2$ behaviour characteristic of point-like particles.
The 5D framework predicts that above the KK/compositeness scale interactions between the SM fields and the pseudo-Goldstone become soft, so that  the form factors are exponentially damped.
The two scales entering the form factors are related to geometrical quantities in 5D:  
\beq
\mkk = {1 \over \int_0^L dy a^{-1}(y)}   \qquad 
f_h  =  {1 \over g_5  \left (\int_0^L dy a^{-2}(y) \right )^{1/2} }  \, .
\eeq 

This paper is organized as follows. 
In \sref{5d} we introduce the formalism that allows us to determine the spectrum of the 5D theory in a background-independent manner. 
We use these results in \sref{ol} to compute the one-loop effective potential for the holographic pseudo-Goldstone boson.  
In \sref{d} we discuss how this formalism can applied for predicting the electroweak observables, such as the Higgs boson mass.

\section{5D framework}
\label{s.5d} 

We are interested in 4D strongly coupled theories, with the global symmetry group $G$ broken to a subgroup $H$ by strong interactions. 
The subgroup $K$ of the surviving global symmetry is weakly gauged by external vector fields (in phenomenological applications, $K$ is typically the SM gauge group).  
The 4D setup contains also chiral fermions coupled to composite operators of the strong sector. 
We describe below a 5D model with analogous properties. 

We consider 5D gauge theories with the  gauge group $G$. 
Four dimensions are flat and non-compact, while the fifth dimension is  an interval, $x_5 \in [0,L]$.
The gravitational background is curved because of  the warp factor $a(x_5)$. 
The line element is    
\beq
\label{e.wb}
ds^2 = a^2(x_5) \eta_{\mu\nu} dx^\mu dx^\nu - dx_5^2 \, .
\eeq  
We fix $a(0) = 1$. 
The choice $a(x_5) = 1$ corresponds to 5D flat space, while $a(x_5) = e^{- k x_5}$ corresponds to a slice of \ads. 
For most of the subsequent discussion we do not specify the warp factor.
We only assume that it is a monotonic and non-increasing function, so that it makes sense to talk about a UV brane at  $x_5 = 0$ and an IR brane at $x_5 = L$, where the value of the warp factor is $a(L) \equiv a_L \leq 1$.

The bulk contains gauge fields $A_M = A_M^\alpha T^\alpha$, where $T^\alpha$ are hermitian generators of the fundamental representation of $G$. 
We also introduce bulk fermions $\psi$, which   
transform under a representation $t^\alpha$ of $G$.  
The 5D action reads  
\beq
S_{\mathrm 5 \mathrm D}= \int d^4 x \int_0^L dx_5 \sqrt{g} \left (
-{1 \over 2 g_5^2} \tr \{ F_{MN}F^{MN} \} 
+  \ov \psi (i \Gamma^N D_N - M)\psi
\right )  \, ,  
\eeq 
where $D_N = \pa_N - i A_N^\alpha t^\alpha$ and $g_5$ is the 5D dimensionful gauge coupling.  

The 5D gauge symmetry is broken down to the subgroup $H \subset G$ on the IR brane and to $K \subset H$ on the UV brane.
To simplify the subsequent discussion and reduce the number of indices we henceforth set\footnote{
Models with $K =  H$ cannot be fully realistic. 
Firstly, in such case the Weinberg angle cannot be made consistent with experiment \cite{grwu}. 
Secondly, consistency with electroweak precision measurements typically requires $H$ to include custodial symmetry.       
However our results can  be easily extended to the realistic case $K = \gsm \subset  H$.} 
$K = H$. 
We divide the generators of $G$ as follows. 
The generators of $H$ are denoted by $T^a$ and $t^a$, while the generators from the coset $G/H$ are denoted by $T^\ha$ and $t^\ha$. 
The commutation relations can be written as  
 \beq
[T^a,T^b] = i f^{abc} T^c \, ,
\quad
[T^a,T^\hab] = i f^{a\hab \hac} T^\hac \, ,
\quad 
[T^\ha,T^\hab] = i f^{\ha\hab c} T^c  \, .
\eeq
The  fermionic representation is divided as $\psi = (\psi^m, \psi^\ham)$, in such a way that  
the generators satisfy $(t^a)_{m\han} =  (t^a)_{\ham n} = 0$ and 
$(t^\ha)_{m n} =  (t^a)_{\ham \han } = 0$.
We impose the boundary conditions 
\beq
\pa_5 A_\mu^a = A_\mu^\ha = A_5^a = \psi^m_R =  \psi_L^\ham  = 0 \, , \qquad x_5 = 0,L \, .
\eeq
One result of such boundary conditions is that the low energy theory contains 4D scalar excitations $h^\ha$, which originate from $A_5^\ha$. 
These scalars have a flat potential at tree level and they can be interpreted as a dual description of the Goldstone bosons arising after spontaneous breaking of the global symmetry in 4D.     
There are  also zero modes  corresponding to $A_\mu^a$, $\psi_L^m$ and $\psi_R^\ham$.
They are interpreted as external fundamental fields mixing with composite states of the 4D strong sector.  
These zero modes are all massless as long as the scalar $h^\ha$ does not develop a vev. 

Our goal is to determine the spectrum of the theory in the presence of the vev 
$h = \la (h^\ha h^\ha)^{1/2} \ra $. 
To this end we rewrite the 5D theory in the KK basis 
\bea
\label{e.kkp} 
A_\mu^a(x,x_5) =   
\sum_n f_n^a(x_5,h) A_{\mu,n}(x)
&\quad&   
A_5^a(x,x_5) = \sum_n {\pa_5 f_n^a(x_5,h) \over m_n(h)} h_n(x)
\nn
A_\mu^\ha(x,x_5) =  \sum_n f_n^\ha(x_5,h) A_{\mu,n}(x)
&\quad&   
A_5^\ha(x,x_5) =  {C_h \over  a^{2}(x_5)} h^\ha(x)  + \sum_n {\pa_5 f_n^\ha(x_5,h) \over m_n(h)} h_{n}(x)
\nn
\psi_{L}^m(x,x_5) = 
\sum_n f_{L,n}^m(x_5,h) \psi_{L,n}(x)
&\quad&
\psi_{R}^m(x,x_5) =  \sum_n f_{R,n}^m(x_5,h) \psi_{R,n}(x) 
\nn
\psi_{L}^\ham(x,x_5) =   \sum_n f_{L,n}^\ham (x_5,h) \psi_{L,n}(x)
&\quad&
\psi_{R}^\ham(x,x_5) = 
\sum_n f_{R,n}^\ham(x_5,h) \psi_{R,n}(x) \, .
\eea
Above we have singled out the scalars $h^\ha(x)$ and the corresponding KK profile. 
The normalization constant should be chosen as  
$C_h = g_5 (\int_0^L a^{-2})^{-1/2}$
so that the scalars are canonically normalized. 
The KK profiles $f_n(x_5,h)$ should be chosen such that the 5D action is rewritten in terms of a tower of 4D fields $A_n(x)$ and $\psi_n(x)$ whose kinetic and mass terms are  diagonal in $n$: 
\bea & \ds 
S_{\mathrm 5 \mathrm D}   =  \int d^4 x \left \{ 
{1 \over 2} (\pa_\mu h^\ha)^2 
+ \sum_n \left (
 -{1 \over 4} [\pa_\mu A_{\nu,n} - \pa_\nu A_{\mu,n}]^2 + {1 \over 2} m_n^2(h) A_{\mu,n}^2
\right )  
\bnl\ds
+ \sum_n \ov \psi_n [\gamma^\mu\pa_\mu - \ti m_n(h)] \psi_n 
+ \dots  \right \} 
\eea
(the dots stand for cubic and quartic interactions and for gauge bosons kinetic mixing with the Goldstones.) 
This is achieved  if the profiles solve the equations of motion in the presence of the scalar vev, and satisfy the boundary conditions  
\beq
\label{e.bc}
\pa_5 f_n^a(x_5,h) = f_n^\ha(x_5,h) = 
f^m_{R,n}(x_5,h) =  f_{L,n}^\ham(x_5,h)    = 0 \, , \qquad x_5 = 0,L \, .
\eeq
The  profiles also satisfy normalization conditions, which are not important in what follows.  

The equations of motion in the presence of the scalar vev are complicated, as they mix Neumann and Dirichlet modes.
However, the 5D gauge invariance relates them to the solutions with $h = 0$.  
For the gauge fields we have  
\beq
\label{e.vrt} 
f^\alpha(x_5,h) T^\alpha = \omega^{-1}(x_5,h) f^\alpha(x_5,0) T^\alpha \omega(x_5,h) ,
\eeq
where $\omega(x_5,h)$ is the gauge transformations that removes the vev of $h^\ha$:
\beq
\omega(x_5,h) = \exp (-i C_h h^\ha T^a \int_0^{x_5} dy \, a^{-2}(y)) .
\eeq 
For $h = 0$ the gauge KK profiles satisfy a simple equation of motion: 
\beq
\label{e.geom} 
\left (\pa_5^2  + 2 {a' \over a} \pa_5 +  {m_n^2 \over a^2} \right ) f_n^\alpha(x_5,0) = 0
\eeq
which is the same for the Dirichlet and the Neumann modes. 
We denote the two independent solutions by $C(x_5,m_n)$ and $S(x_5,m_n)$. 
We choose them such that they  satisfy the initial conditions
$C(0,z) = 1$, $C'(0,z) = 0$, $S(0,z) = 0$, $S'(0,z) = z$.  
These {\it base functions} can be viewed as a warped generalization of the cosine and the sine function 
(in flat 5D $C= \cos(x_5 z)$, $S= \sin(x_5 z)$). 
Then the gauge KK profiles can we written as  
\beq
f_n^a(x_5,0) = C_{n,a} C(x_5,m_n) \qquad  f_n^\ha(x_5,0) = C_{n,\ha} S(x_5,m_n) \,.
\eeq
We can now derive  $f_n^\alpha(x_5,h)$ using \eref{vrt}.
Note that the resulting profiles satisfy the UV boundary conditions \erefn{bc}.   
This is because we have chosen the transformation matrix $\omega$ in such a way that, for $x_5 = 0$, $f_n^\alpha(x_5,0)= f_n^\alpha(x_5,h)$. 
Now, the IR boundary conditions fix the quantization of the KK masses $m_n(h)$. 
After some group algebra we obtain  
\beq
\cos^2 (\lambda_r h/f_h)  C'(L,m_n) S(L,m_n) +  
\sin^2 (\lambda_r h/f_h)  S'(L,m_n) C(L,m_n) = 0 \, ,
\eeq
where the ``Higgs decay constant'' is defined as 
\beq
f_h^2  =  {1 \over g_5^2 \int_0^L dy a^{-2}(y) } 
\eeq
and $(\lambda_r h)^2$ are eigenvalues of the symmetric matrix $\cm$:   
\beq
\label{e.gge}
\cm_{ab} =  h^\ha h^\hab f^{a \ha \hac} f^{b \hab \hac}  \, . 
\eeq
Furthermore, using the Wronskian relation 
\beq 
\label{e.wr}
S'(x_5,z) C(x_5,z) - C'(x_5,z) S(x_5,z) = z \, a^{-2}(x_5)
\eeq 
we can rewrite the quantization condition in a more convenient form: 
\beq
\label{e.gsf} 
1 +  F_{(1)}(m_n^2) \sin^2 \left ( {\lambda_r h \over f_h } \right ) = 0 
\, ,\qquad  \qquad 
F_{(1)}(z^2) = {z  \over a_L^{2} C'(L,z) S(L,z)}  \, .
\eeq 
The set of discrete numbers $\lambda_r$ depends on the gauge groups $G$ and $H$ and $r$ runs over all zero mode gauge bosons. 
For example, for $G = SU(3)$, $H= SU(2) \times U(1)$ from \eref{gge} we find  
$\lambda_1 = 0$ (the photon), $\lambda_2 = \lambda_3 = 1/2$ (the W bosons),   $\lambda_4 = 1$ (the Z boson with a wrong Weinberg angle).\footnote{%
The correct $Z$ boson mass can be adjusted by adding a $U(1)_X$ factor to $G$ and $H$ and breaking one combinations of $U(1)$'s by UV boundary conditions. 
The quantization conditions are very similar in this case, only the form factor for the $Z$ boson is multiplied by a number that depends on the mixing angle between the $U(1)$'s.}

Proceeding analogously with fermionic KK profiles we obtain  
\beq
\label{e.fsf} 
1 +  F_{(1/2)}(m_n^2) \sin^2 \left ( {\lambda_r h \over f_h } \right ) = 0
\,,\qquad \qquad  
F_{(1/2)}(z^2) = - {1 \over S_M(L,z) S_{-M}(L,z)}  \, .
\eeq 
Here $S_M(x_5,z)$ is a solution of the differential equation
\beq
\label{e.feom}
\left [\pa_5^2  + \left ({a' \over a} + 2M \right) \pa_5   + {z^2 \over a^2} \right ] f = 0
\eeq
that satisfies the boundary conditions $S_M(0,z) = 0$, $S_M'(0,z) = z$, and  $(\lambda_r h)^2$ denote eigenvalues  of the matrix: 
\beq
\label{e.fge}
\cm_{mn} = h^\ha h^\hab(t^\ha)_{m\han} (t^\hab)_{\han n} \, . 
\eeq
 
The equations \erefn{gsf} and \erefn{fsf} are the starting point for computing the one-loop effective potential for the scalars $h^\ha$. 

\section{One-loop potential for the pseudo-Goldstone}
\label{s.ol}

\subsection{Coleman--Weinberg potential in 5D} 

The Coleman--Weinberg  formula in 5D KK theories takes the form 
\beq
V = {N \over 2} \sum_n \int {d^4 p \over (2\pi)^4} \log \left (p^2 +  m_n^2 \right ) \, .
\eeq 
Here $N$ is a number of degrees of freedom of a given particle and the sum goes over an infinite KK tower. 
We can regulate the 4D momentum integral using the dimensional regularization, with $d = 4 + \eps$:  
\beq
\label{e.cw}
V = {N \over (4 \pi)^{d/2} \Gamma(d/2)} \sum_n \int  dp p^{d-1} \log \left (1 + {m_n^2 \over p^2} \right ) \, .
\eeq  
In the 5D framework the pseudo-Goldstone boson potential is finite, so the result  does not depend on the way we regularize integrals. 
Now, in dimensional regularization we can evaluate the 4D momentum integral: 
\beq  
V = {N \over (4 \pi)^{d/2}} {\pi \over d \, \Gamma(d/2) \sin(\pi d/2) } \sum_n m_n^d \, .
\eeq
Using  techniques of holomorphic functions, the sum over the KK modes can be traded for an integral. 
To this end we need  a spectral function $\rho(z^2)$, which is holomorphic in the ${\rm Re \, z} >0$ part of the  complex plane and whose zeros on the real axis encode the KK spectrum, $\rho(m_n^2) = 0$. 
Having this, we can rewrite the sum as (see e.g. \cite{odwe}) 
\beq
\sum_n m_n^d  = {d \sin (\pi d/2) \over \pi} \int_{0}^\infty dy  y^{d-1} \log \rho(-y^2) \, .
\eeq 
The Coleman--Weinberg potential thus becomes
\beq
V = {N \over (4 \pi)^{d/2} \Gamma(d/2)}  \int_0^\infty dk k^{d-1} \log \rho(-k^2) \, .
\eeq 
We showed in the previous section that, in the 5D framework, the spectral function encoding the KK spectrum in the presence of the pseudo-Goldstone vev is given by 
\beq
\rho(z^2) = 1 +  F(z^2) \sin^2 \left ( {\lambda_r h \over f_h } \right ) \, ,
\eeq
with some form factor $F(z^2)$ that depends on the 5D gravitational background (for fermions, also on bulk masses).  
At this point we can  safely remove the regulator and set  $d=4$. 
So finally, the scalar potential for the pseudo-Goldstone can be written as 
\beq
\label{e.hcw}
V(h) = \sum_r {N_r \over (4 \pi)^{2}}  
\int_0^\infty dp p^{3} \log \left ( 1 +  F_{r}(-p^2) \sin^2(\lambda_r h/f_h ) \right ) \, . 
\eeq 
The sum now goes over all gauge and fermionic zero modes;
$N_r = +3$ for gauge bosons, $N_r = - 4$ for fermions.
The form factor $F_{(1)}$ for gauge bosons is given by  \eref{gsf},  $F_{(1/2)}$ for fermions is given by  \eref{fsf}.

The scalar potential \erefn{hcw} has the form we would expect in 4D theories with a composite pseudo-Goldstone boson arising as a consequence of spontaneous global symmetry breaking in a strongly interacting sector.
In that set-up the form factor describes an effective interaction vertex between the pseudo-Goldstone and fundamental particles in a low energy theory below the compositeness scale.
The 5D theory predicts the same form on the pseudo-Goldstone potential, which makes the holographic interpretation possible.
The advantage of the 5D framework is that also the form factor $F(z^2)$ is calculable, once we can solve the 5D equations of motion. 
In the remainder of this section we study the shape of the form factors predicted  by 5D theories. 

\subsection{Properties of the form factors}

Even without assuming a specific warp factor $a(x_5)$ and finding explicit solutions to the equations of motion we are able to say a lot about the behaviour of the form factors in various energy regimes.
It turns out that for an arbitrary warp factor the equations of motion can be formally solved below and above the KK scale $\mkk$, where $\mkk$ roughly equals the mass of the lightest resonance. 
  
\ben
\item {\bf Low-energy behaviour.} 
We first investigate the form factors at  small momenta.   
Thus, we consider the equations of motion \erefn{geom} and \erefn{feom} with the $z^2/a^2$  term small enough to be treated perturbatively.  
In this regime the base solutions can be written as   
\bea
\label{e.jysz}
C(x_5,z) &=&  1 -  z^2 \int_0^{x_5} dy \, y \, a^{-2}(y) + \co(z^4)
\nn
S(x_5,z) &=&  z \int_0^{x_5}dy \, a^{-2}(y) + \co(z^3)
\nn
S_M(x_5,z) &=&  z  \int_0^{x_5} dy \, a^{-1}(y)e^{-2 M y}  + \co(z^3) \, .
\eea
The form factors can be  approximated by
\bea
\label{e.ffsz} 
F_{(1)}(-p^2 ) &\approx&  {g^2 f_h^2 \over p^2} \qquad 
g = {g_5\over \sqrt{L}}
\nn
F_{(1/2)}(-p^2) &\approx& {y^2 f_h^2 \over p^2} \qquad
y^2  = {L \int_0^L dy \, a^{-2}(y) \over 
\int_0^L dy \, a^{-1}(y) e^{-2M y} \int_0^L dy \, , a^{-1}(y) e^{2M y}} \, .
\eea 
As could be expected, below the KK scale the form factors exhibit the $1/p^2$ momentum dependence
characteristic of point-like particles. 
From eqs.\erefn{gsf} and \erefn{fsf} the couplings $g$ and $y$ are related to the mass of the light particles: 
\bea
\label{e.lpm} 
m_r^2  &\approx& g^2 f_h^2\sin^2 \left ( {\lambda_r h \over f } \right )  + \co(m_r^4/\mkk^2)\qquad {\rm gauge}
\nn
m_r^2  &\approx& y^2 f_h^2 \sin^2 \left ( {\lambda_r h \over f } \right )  + \co(m_r^4/\mkk^2) \qquad {\rm fermion} .
\eea
\item {\bf Intermediate-energy behaviour.} 
Next, we consider momenta above the KK scale (we will define it shortly).   
In the equations of motion we assume that the  $z^2/a^2$ term is sizable close to  $x_5 \sim L$, while it remains a small perturbation for $x_5 \sim 0$.
Close to the UV brane the base solutions are still well approximated by \eref{jysz}. 
Close to the IR brane we assume that the $\pa_5$ term in the equations of motion can be neglected. 
Under this assumption the base solutions for $x_5 \sim L$ can be written as 
\bea
\label{e.jyiz}
C(x_5,z) &\approx &      a^{-1/2}(x_5) \alpha \cos (z \int_0^{x_5}a^{-1}(y) + \phi_c ) 
\nn
S(x_5,z) &\approx &      a^{-1/2}(x_5) \beta  \sin (z \int_0^{x_5}a^{-1}(y) + \phi_s)  
\nn
S_M(x_5,z) &=&      \beta_M  e^{-M x_5}\sin (z \int_0^{x_5}a^{-1}(y) + \phi_M)  
\eea
This approximation is valid for $x_5$ satisfying $|a'(x_5)| \ll |z|$, $|M| \ll z$.
The amplitude and the phase shifts depend on the detailed behaviour of the solutions in the crossover region and cannot be determined by our general analysis.
They satisfy though the useful relations:  
$\beta \alpha \cos (\phi_c - \phi_s) = 1$
and  $\beta_M \beta_{-M} \cos (\phi_M - \phi_{-M}) = 1$.   
The former follows from the Wronskian\erefn{wr}, 
while the latter from the Crowian: 
\beq  
\label{e.cr} 
a^2(x_5) S_M'(x_5,z) S_{-M}'(x_5,z) +  z^2 S_M(x_5,z) S_{-M}(x_5,z) = z^2 \, .
\eeq

The form factors in the intermediate regime can be approximated by 
\bea 
F_{(1)}(z^2) &\approx& 
- {\cos(\phi_c - \phi_s)  \over 
\sin(\int_0^L a^{-1}(y) z + \phi_c) \sin(\int_0^L a^{-1}(y) z + \phi_s) }
\nn
F_{(1/2)}(z^2) &\approx&  
-{\cos (\phi_M - \phi_{-M}) 
\over 
 \sin (z \int_0^{L}a^{-1}(y) + \phi_{M}) 
  \sin (z \int_0^{L}a^{-1}(y) + \phi_{{-M}}) } \, .
\eea 
It follows that 
\beq
\label{e.ed} 
F_{(1)}(-k^2) \approx F_{(1/2)}(-k^2) \approx e^{-2 k/\mkk}   \, ,
\qquad 
\mkk = {1 \over \int_0^{L} dy a^{-1}(y)} \, . 
\eeq
Above the KK scale the form factors are exponentially damped. 
From eqs.\erefn{gsf} and \erefn{fsf} the KK is  related to the mass of the lightest KK  excitations: 
\beq
m_1 \approx \mkk (\pi - \phi)   \, .
\eeq 

\item {\bf High-energy behaviour.} 
For very high momenta the warping of the 5D space-time should no longer be visible.  
In the equations of motion this  flat-like  regime occurs for large enough $z$,  such that the $\pa_5$ term can be neglected in the whole interval, including $x_5 \sim 0$.  
Then the approximate base solutions are given by: 
\bea
\label{e.jylz}
C(x_5,z) &\approx&      a^{-1/2}(x_5) \cos \left (z \int_0^{x_5}a^{-1}(y) \right ) 
\nn
S(x_5,z) &\approx&      a^{-1/2}(x_5) \sin \left(z \int_0^{x_5}a^{-1}(y) \right) 
\nn 
S_M(x_5,z) &\approx&      e^{-M x_5} \sin \left(z \int_0^{x_5}a^{-1}(y)\right )  \, .
\eea
They are similar to the ones in the intermediate regime, but now  the amplitudes and the phase shifts can be fixed by the boundary conditions at $x_5 = 0$. 
The form factors are given by 
\beq
F_{(1)}(z^2) \approx 
- {1 \over  \sin^2(\int_0^L a^{-1}(y) z ) }
\quad 
F_{(1/2)}(z^2) \approx  
-{1  \over  \sin^2(z \int_0^{L}a^{-1}(y))} \, .
\eeq
For $z^2 = -k^2$ the form factors exhibit an exponential damping, as in \eref{ed}. 
This ensures the convergence of the momentum integral and, in consequence, a finiteness of the pseudo-Goldstone potential at one loop.  
\een
We can easily write a function that interpolates between the $1/k^2$ behaviour at low energy and the exponential damping above the KK scale:  
\beq
\label{e.ffsa} 
F_{(1)}(-p^2) \approx  {g^2 f_h^2 \over \mkk^2 \sinh^2(p/\mkk)} \qquad
F_{(1/2)}(-p^2) \approx  {y^2 f_h^2 \over \mkk^2 \sinh^2(p/\mkk)}  \, .
\eeq
This naive guess works astonishingly well in  many cases. We will see the accuracy of \eref{ffsa} by comparing the approximate with the exact form factors derived by solving the equations of motion in  specific backgrounds.
The approximation of the fermionic warp factor is less accurate for large bulk masses.
This could be improved, but large bulk masses are not important in practice, as they lead to suppressed contributions to the pseudo-Goldstone potential.

\subsection{Exact form factors}

Below, we give a few examples of solutions to the equations of motion  in specific backgrounds and we calculate the corresponding form factors. 

\subsubsection{Flat spacetime} 

For the 5D Minkowski spacetime we set $a(x_5) = 1$. 
As $a(L)/a(0) = 1$ this background does not have a holographic interpretation, but it is the simplest one to perform 5D computations.  

The base functions solving eqs.\erefn{geom} and \erefn{feom} are given by   
\beq
C(x_5,z) = \cos (z x_5) \qquad   S(x_5,z) = \sin(z x_5) \qquad 
S_M(x_5,z) =  {z e^{-M x_5} \over \sqrt{z^2 -M^2}}  \sin(\sqrt{z^2 -M^2} x_5).
\eeq
Inserting this into eqs.\erefn{gsf} and \erefn{fsf} we obtain \cite{flat}
\beq
F_{(1)}(-p^2) = {1 \over \sinh^2 (p \, L)}
\qquad 
F_{(1/2)}(-p^2) = {p^2 + M^2 \over p^2} {1 \over \sin^2 (\sqrt{p^2+M^2} L)} \, .
\eeq 
In this case our approximate expression \erefn{ffsa} for the gauge form factor, with $g f_h = \mkk = 1/L$,  is equal to the true. 
The fermionic approximation is fine for $M L \ll 1$.

\subsubsection{$\bf \mathrm{AdS}_5$ spacetime} 

We take $a(x_5) = e^{-k x_5}$ and, customarily, we parametrize the bulk fermion mass as  $M = c k$.  
This is the Randall--Sundrum spacetime \cite{rasu}, which arises when the background is shaped by 5D gravity with a negative cosmological constant, $V = -6 k^2$. 
5D theories in such background are conjectured to be dual to 4D theories, with a  conformally symmetric strong sector \cite{adscftrs}.
The parameter $c$ is related to an anomalous dimension of the fermionic composite operators, to which the fundamental fermions are coupled \cite{copo}.

Solutions to eqs.\erefn{geom} and \erefn{feom} can be expressed in terms of Bessel functions \cite{po}. 
The base functions with prescribed UV boundary conditions are given by the following expressions:
\bea
C(x_5,z) &=& {\pi z \over 2 k} a^{-1}(x_5) \left [
 Y_0 \left ( {z \over k } \right )      J_1 \left ( {z \over k a(x_5)} \right ) 
- J_0 \left ( {z \over k} \right )     Y_1 \left ( {z \over k a(x_5)} \right )  
\right ]
\nn
S(x_5,z) &=&  {\pi z  \over 2 k} a^{-1}(x_5) \left [
 J_1 \left ( {z \over k } \right )      Y_1 \left ( {z \over k a(x_5)} \right ) 
- Y_1 \left ( {z \over k} \right )     J_1 \left ( {z \over k a(x_5)} \right )  
\right ]
\nn
S_M(x_5,z) &=&  {\pi z  \over 2 k} a^{c-{1 \over 2} }(x_5) \left [
 J_{{1 \over 2}-c} \left ( {z \over k } \right )      Y_{{1\over 2}-c} \left ( {z \over k a(x_5)} \right ) 
- Y_{{1\over 2}-c } \left ( {z \over k } \right )     J_{{1\over 2}-c}  \left ( {z \over k a(x_5)} \right )  
\right ].
\nn
\eea
The form factors are given by \cite{odwe,honosa} 
\bea & \ds
\label{e.gffrs} 
F_{(1)}(-p^2) =   {a_L  k^2  \over p^2}  
\nl \ds {
1 \over
\left [ I_0 \left ( {p \over k } \right ) K_0 \left ( {p \over k a_L} \right )  
-  K_0 \left ( {p \over k } \right )   I_0 \left ( {p \over k a_L} \right ) \right ] 
\left [ I_1 \left ( {p \over k} \right ) K_1 \left ( {p \over k a_L} \right )  
-  K_1 \left ( {p \over k } \right )   I_1 \left ( {p \over k a_L} \right ) \right ] 
} 
\eea
\bea & \ds
\label{e.fffrs} 
F_{(1/2)}(-p^2) =   {a_L k^2  \over p^2}  
\nl \ds { 1 \over
\left [ I_{{1\over 2}-c} \left ( {p \over k } \right ) K_{{1\over 2}-c} \left ( {p \over k a_L} \right )  
-  K_{{1\over 2}-c} \left ( {p \over k } \right )   I_{{1\over 2}-c} \left ( {p \over k a_L} \right ) \right ]
\left [ I_{{1\over 2}+c} \left ( {p \over k} \right ) K_{{1 \over 2}+c} \left ( {p \over k a_L} \right )  
-  K_{{1 \over 2}+c} \left ( {p \over k} \right )   I_{{1 \over 2}+c} \left ( {p \over k a_L} \right ) \right ] 
}.
\nn
\eea
These form factors can be approximated by those in  \eref{ffsa} with    
\beq
(g f_h)^2 =  {2 k^2 \over k L(a_L^{-2} - 1)}
\qquad \mkk = {k \over (a_L^{-1} - 1)} 
\qquad 
(y f_h)^2  =  {(2 c+ 1)(-2c+1) k^2 \over (a_L^{2c-1} - 1)(a_L^{-2c-1} - 1)}  \, . 
\eeq
The accurateness of these approximations is visualized in \fref{rsa}. 
The approximate of the gauge form factor precisely traces the true one. 
For fermions the approximate is even better for $|c| \leq 1/2$,  while it  deviates for  $|c| \gg 1/2$. 

\begin{figure}[b]
\begin{center}
\includegraphics[width=0.45\textwidth]{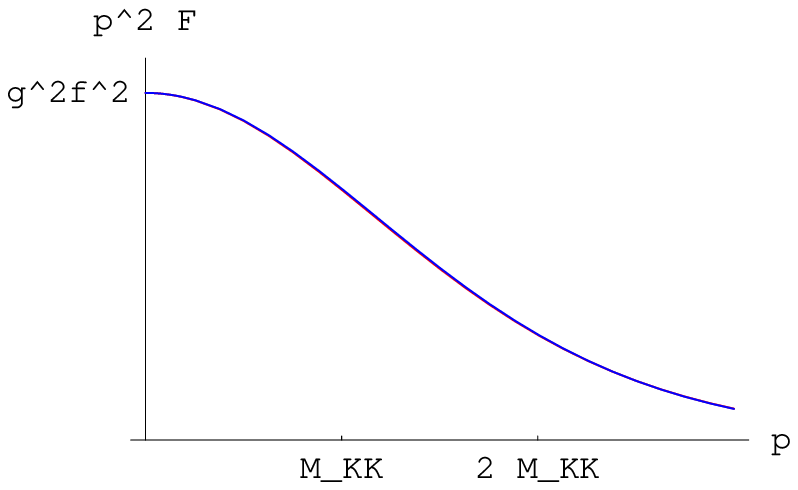}
\includegraphics[width=0.45\textwidth]{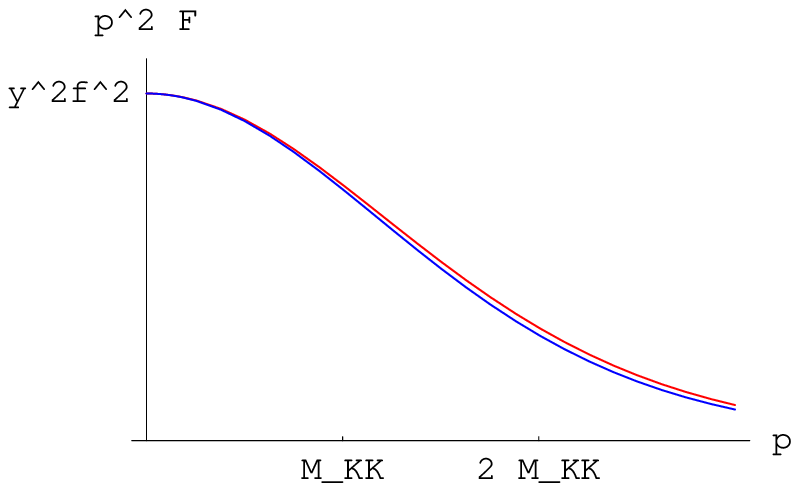}
\caption{\footnotesize 
Left panel: the true (red) and the approximate (blue) gauge form factor in \ads\, for $a_L = 10^{-3}$. 
If you see only one line, it's because the approximate is so successful ;-).
Right panel: the same for the fermionic form factor, with $c = 0.7$.   
}   
\label{f.rsa} 
\end{center}
\end{figure}

\subsubsection{Dilatonic spacetime}

One more example where the gauge equation of motion can be solved (the fermionic one is more resistant)  
is the 5D spacetime with the warp factor varying as a power of $x_5$:   
\beq
\label{e.plwf}
a(x_5) = \left ( 1 - {k x_5 \over \gamma - 1} \right )^\gamma \, .
\eeq
For definiteness we assume   $\gamma > 1$, so that $a_L < 1$ for positive $k$.
Of course, we need $k L < \gamma  - 1$ to avoid a naked singularity.  
In the conformal limit $\gamma \to \infty$ we recover the exponential warp factor of \ads.
For finite $\gamma$ the 5D model is dual to 4D strongly coupled theories with conformal symmetry explicitly broken by a coupling $\sim 1/\gamma$ that logarithmically increases towards IR. 

This power-law  warp factor arises as a solution in 5D gravity--dilaton theories with the dilaton potential  
\beq
V(\phi) = - 6 k^2 {\gamma (\gamma - 1/4) \over (\gamma - 1)^2} e^{\pm {2 \over \sqrt{3 \gamma}}(\phi - \phi_0)}
\eeq
and appropriate boundary potentials.
The well-known example where such background appears is the Horava--Witten model \cite{howi} compactified to 5D \cite{luovst}, in which case $\gamma = 1/6$.  

Solutions to \eref{geom} can be written, much as in  the \ads\, case, in terms of the Bessel functions, 
\bea
C(x_5,z) & = &   {\pi z \over 2 k} a^{-1 + {1 \over 2\gamma} } \left [ 
Y_{1 \over 2\gamma-2}  \left ({z \over k } \right ) 
J_{2 \gamma-1 \over 2\gamma-2}  \left ({z \over k a^{1 - {1 \over \gamma}}} \right ) 
- J_{1 \over 2\gamma-2}  \left ({z \over k } \right ) 
Y_{2 \gamma-1 \over 2\gamma-2}  \left ({z \over k a^{1 - {1 \over \gamma}}} \right ) \right ]
\nn
S(x_5,z) & = & {\pi z \over 2 k}
a^{-1 + {1 \over 2\gamma}} \left [ 
J_{2 \gamma-1 \over 2\gamma-2}  \left ({z \over k } \right ) 
Y_{2 \gamma-1 \over 2\gamma-2}  \left ({z \over k a^{1 - {1 \over \gamma}}} \right ) 
- Y_{ 2 \gamma-1 \over 2\gamma-2}  \left ({z \over k } \right ) 
J_{2 \gamma-1 \over 2\gamma-2}  \left ({z \over k a^{1 - {1 \over \gamma}}} \right ) \right ]
\nn
\eea
The gauge form factor is readily calculated: 
\bea &
\label{e.gsfpe} 
\ds F_{(1)}(-p^2) =   { a_L^{1- {1 \over \gamma}} k^2 \over p^2} 
\nl \ds  \cdot {1 
\over
\left [ I_{1 \over 2\gamma-2} \left ( {p \over k } \right ) K_{1 \over 2\gamma-2} \left ( {p \over k a_L^{1-{1 \over \gamma}}} \right )  
-  K_{1 \over 2\gamma-2} \left ( {p \over k} \right )   I_{1 \over 2\gamma-2} \left ( {p \over k a_L^{1-{1\over \gamma} }} \right ) \right ]
}
\cdot
\nl \ds \cdot
{1 \over  
\left [ I_{2 \gamma-1 \over 2\gamma-2} \left ( {p \over k} \right ) I_{2 \gamma-1 \over 2\gamma-2} \left ( {p \over k a_L^{1-{1\over \gamma}}} \right )  
-  K_{2 \gamma-1 \over 2\gamma-2} \left ( {p \over k } \right )   I_{2 \gamma-1 \over 2\gamma-2} \left ( {p \over k a_L^{1-{1\over \gamma}}} \right ) \right ] 
} 
\eea 
The approximate expression is that of \eref{ffsa} with    
\beq
(g f_h)^2 =  {(2\gamma- 1)  k^2 \over (\gamma-1) k L(a_L^{-2+{1/\gamma}} - 1)}
\qquad \mkk = {k \over a_L^{-1+{1/\gamma}} - 1} \, .
\eeq
As can be seen from  \fref{pa}, the approximate expression  works  very well as long as $\gamma$ is sufficiently larger than $1$.  

\begin{figure}[b]
\begin{center}
\includegraphics[width=0.45\textwidth]{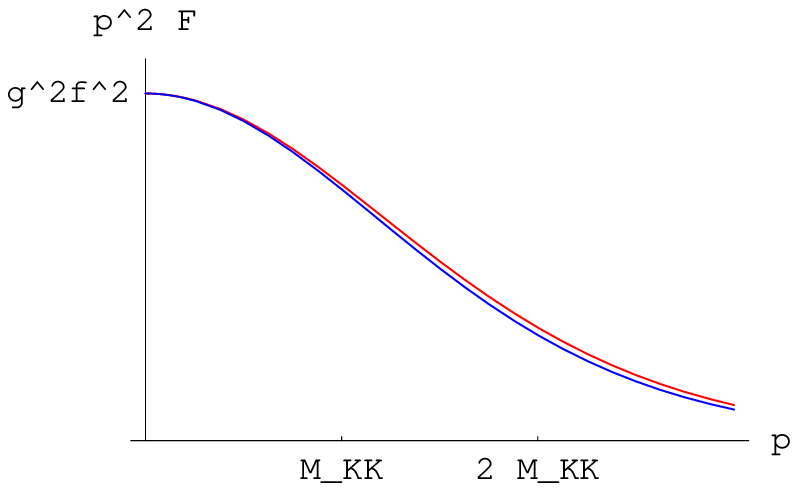}
\includegraphics[width=0.45\textwidth]{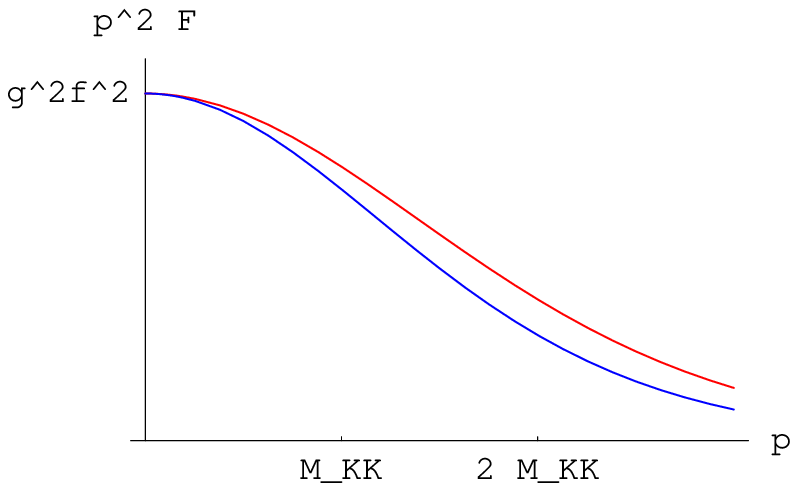}
\caption{\footnotesize 
Left panel: the true (red) and the approximate (blue) gauge form factor in the dilatonic spacetime, for $a_L = 10^{-3}$ and $\gamma = 4$. 
Right panel: the same for $\gamma = 1.5$.   
}   
\label{f.pa} 
\end{center}
\end{figure}

\section{Discussion}
\label{s.d}

We presented a simple algorithm for computing the form factors entering the pseudo-Goldstone potential.  
The methods developed here clarify the existing computations in the flat and \ads\, spacetimes, and facilitate extensions to other 5D backgrounds.  
Our results should be relevant not only for electroweak breaking, but also for the so-called natural inflation \cite{natural}, which also makes use of the Goldstone mechanism.     

Furthermore, we have argued that the form factors obtained in the 5D framework can often be  adequately approximated by \eref{ffsa}.
Consequently, the pseudo-Goldstone potential can be approximated by: 
\beq
\label{e.hcwa}
V(h) \approx \sum_r {N_r \over (4 \pi)^{2}}  
\int_0^\infty dp p^{3} \log \left ( 1 +  
{g_r^2 f_h^2 \over \mkk^2} {\sin^2(\lambda_r h/f_h) \over \sinh^2(p/\mkk) }  \right ) \, ,
\eeq 
where the sum goes over all zero modes.
Recall that the two scales entering the potential are simply related to 5D geometrical quantities:  
\beq
\label{e.scales}
\mkk = {1 \over \int_0^L dy a^{-1}(y)}  \, , \qquad 
f_h^2  =  {1 \over g^2 L \int_0^L dy a^{-2}(y) }  \, .
 \eeq 

We conclude with a simple example showing how our results can be applied in phenomenological situations. 
In the following  we identify the pseudo-Goldstone boson $h^\ha$ with the Higgs field responsible for electroweak symmetry breaking of the SM.  
We study how various observables like the electroweak scale or the Higgs mass depend on the two scales defined in \eref{scales}. 
We assume here that the Higgs potential \erefn{hcwa} is dominated  by a contribution of only one low energy particle species, the top quark. 
Then we can write the Higgs potential as 
\beq
\label{e.hcwt}
V(h) \approx  - {3 \over 4 \pi^{2}} \mkk^4   
\int_0^\infty dp p^{3} \log \left ( 1 +  
{y_t^2 f_h^2 \over \mkk^2} {\sin^2(\lambda_t h/f_h) \over \sinh^2(p) }  \right ) \, .
\eeq 
At the origin, $V''(0) < 0$, so the Higgs develops a vev and electroweak symmetry is broken.  
The minimum occurs at $\sin(\lambda_t h/f_h) = 1$, thus 
$\la h \ra = {\pi f_h /2 \lambda_t}$.
Consequently, the top mass and the $W$ mass are given by  
\beq
m_t = y_t f_h \, , \qquad m_W = g f_h \sin(\lambda_W \la h \ra /f_h) \, . 
\eeq
The relation between the Higgs decay constant and the electroweak scale depends on group representations via $\lambda_W$, but typically $m_W \sim g f_h$, where $g$ is the $SU(2)_W$ gauge coupling constant.
Thus, in this simple example it is not possible to separate the electroweak scale from the global symmetry breaking scale. 
Therefore, for  the model to be phenomenologically relevant, we need to separate the KK scale from the electroweak scale $m_W \sim g f_h \ll \mkk$. 
From  \eref{scales} the ratio is given by 
\beq
{\mkk \over g f_h}  = 
{\left ( L \int_0^L dy a^{-2}(y) \right )^{1/2} \over \int_0^L dy a^{-1}(y)}  \, .
\eeq
For flat 5D spacetime this ratio is $1$, which is of course unacceptable. 
But, in general, warping leads to a desired scale separation.
For the Randall--Sundrum model one finds that the scale separation is logarithmically enhanced \cite{odwe, agcopo,honosa}
\beq
\mkk/g f_h \approx \sqrt{kL/2} \sim \log^{1/2}(\mpl/\mkk) \, . 
\eeq
For the power-law warp factor \erefn{plwf} we find
\beq
\mkk/g f_h \approx  \sqrt{k L (\gamma-1)/(2 \gamma-1) } a_L^{-1/2\gamma}  \, 
\eeq
and even  larger scale separation can be obtained for $a_L \ll 1$ and $\gamma$ away from the conformal limit.  

Computing the second derivative of \eref{hcwt} at the minimum we find the Higgs boson mass
\beq
m_h^2 = {9 \over 4 \pi^2} \lambda_t^2  m_t^2  {\mkk^2 \over f_h^2} \, \alpha(m_t^2/\mkk^2)  \, ,
\eeq
where $\alpha(x) = (2/3)\int_0^\infty t^3 (\sinh^2 t + x)^{-1}$ is a slowly varying function $\sim 1$; 
$\alpha(0) = \zeta(3)$, $\alpha(1) = (3/4) \zeta(3)$. 
The Higgs mass is suppressed with respect to the electroweak scale by a loop factor, but it is enhanced by 
the ratio $\mkk/f_h$. 
Thus, enough separation between the KK and the electroweak scale automatically ensures that the Higgs boson mass is consistent with current experimental bounds.
This conclusion does not depend whether the 5D background is conformal, as in the Randall--Sundrum model, or if it is of a more general warped form.

This example illustrates how the formalism we developed clarifies the dependence of the observables in the electroweak sector on the scales describing the 5D model or its holographic 4D dual. 
Of course, a full-fledged phenomenological analysis requires computing corrections to electroweak precision observables. 
This is left for future studies.

\section*{Acknowledgements}

I would like to thank Ricardo Rattazzi and Claudio Scrucca  for nice discussions.

I am partially supported by the European Community Contract MRTN-CT-2004-503369 for the years 2004--2008 
and by  the MEiN grant 1 P03B 099 29 for the years 2005--2007.




\begin{thebibliography}{99}


\bibitem{georgi}
  D.~B.~Kaplan and H.~Georgi,
  ``SU(2) X U(1) breaking by vacuum misalignment,''
  Phys.\ Lett.\ B {\bf 136}, 183 (1984).
  M.~J.~Dugan, H.~Georgi and D.~B.~Kaplan,
  ``Anatomy Of A Composite Higgs Model,''
  Nucl.\ Phys.\ B {\bf 254}, 299 (1985).
\bibitem{adscft}
 J.~M.~Maldacena,
  ``The large N limit of superconformal field theories and supergravity,''
  Adv.\ Theor.\ Math.\ Phys.\  {\bf 2} (1998) 231
  [Int.\ J.\ Theor.\ Phys.\  {\bf 38} (1999) 1113]
  [arXiv:hep-th/9711200].

\bibitem{adscftrs}
  N.~Arkani-Hamed, M.~Porrati and L.~Randall,
  ``Holography and phenomenology,''
  JHEP {\bf 0108} (2001) 017
  [arXiv:hep-th/0012148].
  R.~Rattazzi and A.~Zaffaroni,
  ``Comments on the holographic picture of the Randall-Sundrum model,''
  JHEP {\bf 0104} (2001) 021
  [arXiv:hep-th/0012248].
  M.~P\'erez-Victoria,
  ``Randall-Sundrum models and the regularized AdS/CFT correspondence,''
  JHEP {\bf 0105} (2001) 064
  [arXiv:hep-th/0105048].
  T.~Gherghetta,
  ``Les Houches lectures on warped models and holography,''
  arXiv:hep-ph/0601213.

\bibitem{ho}
  Y.~Hosotani,
  ``Dynamical mass generation by compact extra dimensions,''
  Phys.\ Lett.\ B {\bf 126}, 309 (1983).
  Y.~Hosotani,
  ``Dynamics of nonintegrable phases and gauge symmetry breaking,''
  Annals Phys.\  {\bf 190}, 233 (1989).
  H.~Hatanaka, T.~Inami and C.~S.~Lim,
  ``The gauge hierarchy problem and higher dimensional gauge theories,''
  Mod.\ Phys.\ Lett.\ A {\bf 13}, 2601 (1998)
  [arXiv:hep-th/9805067].



\bibitem{flat}
  I.~Antoniadis, K.~Benakli and M.~Quiros,
  ``Finite Higgs mass without supersymmetry,''
  New J.\ Phys.\  {\bf 3}, 20 (2001)
  [arXiv:hep-th/0108005].
  M.~Kubo, C.~S.~Lim and H.~Yamashita,
   ``The Hosotani mechanism in bulk gauge theories with an orbifold extra  space S(1)/Z(2),''
  Mod.\ Phys.\ Lett.\ A {\bf 17}, 2249 (2002)
  [arXiv:hep-ph/0111327].
  L.~J.~Hall, Y.~Nomura and D.~R.~Smith,
  ``Gauge-Higgs unification in higher dimensions,''
  Nucl.\ Phys.\ B {\bf 639}, 307 (2002)
  [arXiv:hep-ph/0107331].
  G.~Burdman and Y.~Nomura,
  Nucl.\ Phys.\ B {\bf 656}, 3 (2003)
  [arXiv:hep-ph/0210257].
  C.~Csaki, C.~Grojean and H.~Murayama,
  ``Standard model Higgs from higher dimensional gauge fields,''
  Phys.\ Rev.\ D {\bf 67}, 085012 (2003)
  [arXiv:hep-ph/0210133].
  C.~A.~Scrucca, M.~Serone and L.~Silvestrini,
  ``Electroweak symmetry breaking and fermion masses from extra dimensions,''
  Nucl.\ Phys.\ B {\bf 669}, 128 (2003)
  [arXiv:hep-ph/0304220].
  C.~A.~Scrucca, M.~Serone, L.~Silvestrini and A.~Wulzer,
  ``Gauge-Higgs unification in orbifold models,''
  JHEP {\bf 0402} (2004) 049
  [arXiv:hep-th/0312267].
  Y.~Hosotani, S.~Noda and K.~Takenaga,
  ``Dynamical gauge-Higgs unification in the electroweak theory,''
  Phys.\ Lett.\ B {\bf 607} (2005) 276
  [arXiv:hep-ph/0410193].
  G.~Cacciapaglia, C.~Csaki and S.~C.~Park,
  ``Fully radiative electroweak symmetry breaking,''
  JHEP {\bf 0603} (2006) 099
  [arXiv:hep-ph/0510366].
  G.~Panico, M.~Serone and A.~Wulzer,
  ``A model of electroweak symmetry breaking from a fifth dimension,''
  Nucl.\ Phys.\ B {\bf 739} (2006) 186
  [arXiv:hep-ph/0510373].

\bibitem{conopo}
  R.~Contino, Y.~Nomura and A.~Pomarol,
  Nucl.\ Phys.\ B {\bf 671} (2003) 148
  [arXiv:hep-ph/0306259].
\bibitem{odwe}
  K.~y.~Oda and A.~Weiler,
   ``Wilson lines in warped space: dynamical symmetry breaking and restoration,''
  Phys.\ Lett.\ B {\bf 606} (2005) 408
  [arXiv:hep-ph/0410061].
 \bibitem{agcopo}
  K.~Agashe, R.~Contino and A.~Pomarol,
  ``The minimal composite Higgs model,''
  Nucl.\ Phys.\ B {\bf 719} (2005) 165
  [arXiv:hep-ph/0412089].
\bibitem{homa}
  Y.~Hosotani and M.~Mabe,
   ``Higgs boson mass and electroweak-gravity hierarchy from dynamical gauge-Higgs unification in the warped spacetime,''
  Phys.\ Lett.\ B {\bf 615} (2005) 257
  [arXiv:hep-ph/0503020].
\bibitem{honosa}
  Y.~Hosotani, S.~Noda, Y.~Sakamura and S.~Shimasaki,
   ``Gauge-Higgs unification and quark-lepton phenomenology in the warped spacetime,''
  Phys.\ Rev.\ D {\bf 73} (2006) 096006
  [arXiv:hep-ph/0601241].

 \bibitem{agco}
  K.~Agashe and R.~Contino,
``The minimal composite Higgs model and electroweak precision tests,''
  Nucl.\ Phys.\ B {\bf 742} (2006) 59
  [arXiv:hep-ph/0510164].
  M.~Carena, E.~Ponton, J.~Santiago and C.~E.~M.~Wagner,
  ``Light Kaluza-Klein states in Randall-Sundrum models with custodial SU(2),''
  arXiv:hep-ph/0607106.


\bibitem{gowi}
  W.~D.~Goldberger and M.~B.~Wise,
  ``Modulus stabilization with bulk fields,''
  Phys.\ Rev.\ Lett.\  {\bf 83}, 4922 (1999)
  [arXiv:hep-ph/9907447].
  O.~DeWolfe, D.~Z.~Freedman, S.~S.~Gubser and A.~Karch,
  ``Modeling the fifth dimension with scalars and gravity,''
  Phys.\ Rev.\ D {\bf 62} (2000) 046008
  [arXiv:hep-th/9909134].
  
 
 


\bibitem{grwu}
  B.~Grzadkowski and J.~Wudka,
  ``5-dimensional difficulties of gauge-Higgs unifications,''
  arXiv:hep-ph/0604225.

       
\bibitem{rasu}
L.~Randall and R.~Sundrum,
``A large mass hierarchy from a small extra dimension,''
Phys.\ Rev.\ Lett.\  {\bf 83} (1999) 3370
[arXiv:hep-ph/9905221].

\bibitem{copo}
  R.~Contino and A.~Pomarol,
  ``Holography for fermions,''
  JHEP {\bf 0411} (2004) 058
  [arXiv:hep-th/0406257].
   
\bibitem{po}
  A.~Pomarol,
  ``Gauge bosons in a five-dimensional theory with localized gravity,''
  Phys.\ Lett.\ B {\bf 486} (2000) 153
  [arXiv:hep-ph/9911294].

\bibitem{howi}
  P.~Horava and E.~Witten,
  ``Eleven-dimensional supergravity on a manifold with boundary,''
  Nucl.\ Phys.\ B {\bf 475} (1996) 94
  [arXiv:hep-th/9603142].

\bibitem{luovst}
  A.~Lukas, B.~A.~Ovrut, K.~S.~Stelle and D.~Waldram,
  ``Heterotic M-theory in five dimensions,''
  Nucl.\ Phys.\ B {\bf 552} (1999) 246
  [arXiv:hep-th/9806051].

\bibitem{natural}
  K.~Freese, J.~A.~Frieman and A.~V.~Olinto,
  ``Natural inflation with pseudo - Nambu-Goldstone bosons,''
  Phys.\ Rev.\ Lett.\  {\bf 65} (1990) 3233.
  N.~Arkani-Hamed, H.~C.~Cheng, P.~Creminelli and L.~Randall,
  ``Extranatural inflation,''
  Phys.\ Rev.\ Lett.\  {\bf 90} (2003) 221302
  [arXiv:hep-th/0301218].
             
\end{thebibliography}
\end{document}